\documentclass[pre,aps,amsmath,amssymb,superscriptaddress]{revtex4}
\usepackage{mathrsfs}
\usepackage{amssymb}
\usepackage{amsmath}
\usepackage{bbm}
\pdfoutput=1

\usepackage{graphicx}
\usepackage{dcolumn}
\usepackage{bm}
\newcommand{\bra}[1]{\left(#1\right)}

\newcommand{\brc}[1]{\left \langle#1\right \rangle}
\newcommand{\bre}[1]{\left\{#1\right\}}

\newcommand{\be}{\begin{equation}}
\newcommand{\ee}{\end{equation}}
\newcommand{\bea}{\begin{eqnarray}}
\newcommand{\eea}{\end{eqnarray}}

\newcommand{\NMI}{{\rm NMI}}
\newcommand{\rNMI}{{\rm rNMI}}

\newcommand{\pin}{p_{\rm in}}
\newcommand{\pout}{p_{\rm out}}
\newcommand{\eps}{\epsilon}

\begin{document}
\title{Evaluating accuracy of community detection using the relative normalized mutual information}
	\author{Pan Zhang}
\affiliation{State Key Laboratory of Theoretical Physics, Institute of Theoretical Physics, Chinese Academy of Sciences, Beijing 100190, China}
\affiliation{Santa Fe Institute, Santa Fe, New Mexico 87501, USA}
\email{pan@santafe.edu}
\begin{abstract}
	The Normalized Mutual Information (NMI) has been widely used to evaluate the accuracy 
	of community detection algorithms. However in this article we show that the NMI 
	is seriously affected by 
	systematic errors due to finite size of networks, and may give a wrong estimate of 
	performance of algorithms in some cases. We give a simple 
	theory to the finite-size effect of NMI and test our theory numerically. 
	Then we propose a 
	new metric for the accuracy of community detection, 
	namely the relative Normalized Mutual Information (rNMI), which considers statistical significance
	of the NMI by comparing it with the expected NMI of random partitions.
	Our numerical experiments show that the rNMI overcomes the finite-size effect of the NMI.
\end{abstract}
\maketitle

Detection of community structures, which asks to group 
nodes in a network into groups, 
is a key problem in network science, 
computer science, sociology and biology. Many algorithms 
have been proposed for this problem, see ref. \cite{Fortunato2010} for a review. 
However on a given network, different algorithms usually give different results. Thus 
evaluating performance of these algorithms and finding the best ones are
of great importance.

Usually the evaluations are performed on benchmark networks each of which
has a reference partition. These benchmarks include
networks generated by generative models, like Stochastic Block Model \cite{holland1983stochastic} and LFR model 
\cite{lancichinetti2008benchmark}, with
a planted partition as the reference partition; and some real-world networks, like 
the famous Karate Club network \cite{Zachary1977} and the Political Blog network 
\cite{Adamic2005}, with 
a partition annotated by domain experts as the reference partition. 
The accuracy of a community detection algorithm 
is usually represented using similarity between the reference partition and 
partition found by the algorithm --- the larger similarity, the better performance the algorithm has on 
the benchmark.

Without losing generality, in what follows we call the reference 
partition $A$ and the detected partition $B$, and our task is to study
the measure of similarity between partition $A$ and partition $B$.

When the number of groups are identical, $q_A=q_B=q$, 
the similarity can be 
easily defined by the overlap, which is the number of identical group labels in $A$ and $B$ maximized
over all possible permutations: 
\begin{equation}
	O(A,B)=	\max_{\pi}\left( \frac{1}{n} \sum_{i=1}^n \delta_{A_i , \pi (B_i)} -
          \frac{1}{q}\right),
\end{equation}
where $n$ is number of nodes, $\delta$ is the Kronecker delta function and $\pi$ ranges over all permutations of $q$ groups.

However we can see that the overlap is non-zero even if partition $B$ is 
a random partition: there are
roughly $\frac{n}{q}$ identical labels in two partitions if labels are distributed 
randomly and uniformly. 
One way to refine it is to normalize the overlap 
to scale from $0$ to $1$ \cite{Decelle2011,Krzakala2013}:
\begin{equation}\label{eq:ovl}
	{{\mathcal{O}}(A,B)}=	\max_{\pi}\left( \frac{1}{n} \sum_{i=1}^n \delta_{A_i , \pi (B_i)} -
          \frac{1}{q}\right) \left/ \left( 1 -  \frac{1}{q} \right) \right. \, .
\end{equation}

However despite its simplicity, using overlap as the similarity has two problems: first, when number of 
groups $q$ is large, maximizing overlap over $q!$ permutations is difficult;
second, when number of partitions, $q_A$ and $q_B$, 
are not identical, the overlap is ill-defined.

Another well-accepted measure of similarity 
is the 
Normalized Mutual Information (NMI) \cite{danon2005comparing,ana2003robust},
which is well-defined even when $q_A\neq q_B$. Many studies use 
NMI to evaluate their algorithms or to compare different algorithms 
\cite{lancichinetti2009community,Fortunato2010}.
To define NMI we need to approximate the marginal probability of a 
randomly selected node being in group $a$ and $b$ by $P_A(a)=\frac{n_a}{n}$
and $P_B(b)=\frac{n_b}{n}$, where 
$n_a$ and $n_b$ denote group size of $A$ and $B$.

We know that the spirit of Mutual Information is to compute the dependence of these two distributions,
by computing Kullback-Leibler (KL) distance between joint distribution $P_{AB}(a,b)$ and the product of 
two marginal distributions $P_A(a)P_B(b)$:
\begin{equation}\label{eq:I1}
	I_{AB}(P_A,P_B)=\sum_{a=1}^{q_a}\sum_{b=1}^{q_b}P_{AB}(a,b)\log\frac{P_{AB}(a,b)}{P_A(a)P_B(b)}.
\end{equation}
Due to the property of KL distance, this quantity is non-negative. 
$I_{AB}(P_A,P_B)=0$ implies that $P_A$ and $P_B$ are independent, 
that detected partition has nothing to do with the ground-true partition.
And if $I_{AB}(P_A,P_B)$ is much larger than zero, 
the detected partition and the ground true partition are similar.
In practice the joint distribution can also be approximated by frequencies
$$P_{AB}(a,b)=\frac{n_{ab}}{n},$$
where $n_{ab}$ is number of nodes that both in group $a$ of partition $A$ and in group
$b$ of partition $B$.
Eq.\eqref{eq:I1} can be written as
\begin{equation}\label{eq:I2}
	I(P_A,P_B)=H(P_A)+H(P_B)-H(P_{AB}),
\end{equation}
where $$H(P_A)=-\sum_aP_A(a)\log P_A(a)$$ is the Shannon entropy of 
distribution $P_A$,
and $H(P_{AB})$ is the entropy of the joint distribution $P_{AB}$.
Note that using conditional distribution $P_{A|B}(a|b)=P_{AB}(a,b)/P_B(b)$, one can 
rewrite Eq.\eqref{eq:I1} as
\begin{equation}\label{eq:I3}
	I(P_A,P_B)=H(P_A)-H(P_{A|B}),
\end{equation}
which has a interpretation that amount of information (surprise) gained on distribution
$P_A$ after known $B$. If this information gain is $0$, it means 
knowledge of $B$ does not give any information about $A$, then two partitions has 
nothing to do with each other.
Obviously the larger $I_{AB}(P_A,P_B)$, the more similar two partitions are. However 
this is still not a ideal metric for evaluating community detection algorithms since
it is not normalized. As proposed in \cite{danon2005comparing}, one way to normalize it is to choose normalization as 
$H(P_A)+H(P_B)$, and the Normalized Mutual Information is written as
\begin{equation}\label{eq:nmi}
	\NMI(P_A,P_B)=\frac{2I(P_A,P_B)}{H(P_A)+H(P_B)}.
\end{equation}
Since $H(P_{AB})\leq H(P_A)+H(P_B)$, $\NMI(P_A,P_B)$ is bounded below 
by $0$.
Also note that $H(P_{AB})= H(P_A)=H(P_B)$ when $A$ and $B$ are identical, 
which means in this case $\NMI(P_A,P_B)=1$.

\begin{figure}  
	\centering
    \includegraphics[width=0.45\columnwidth]{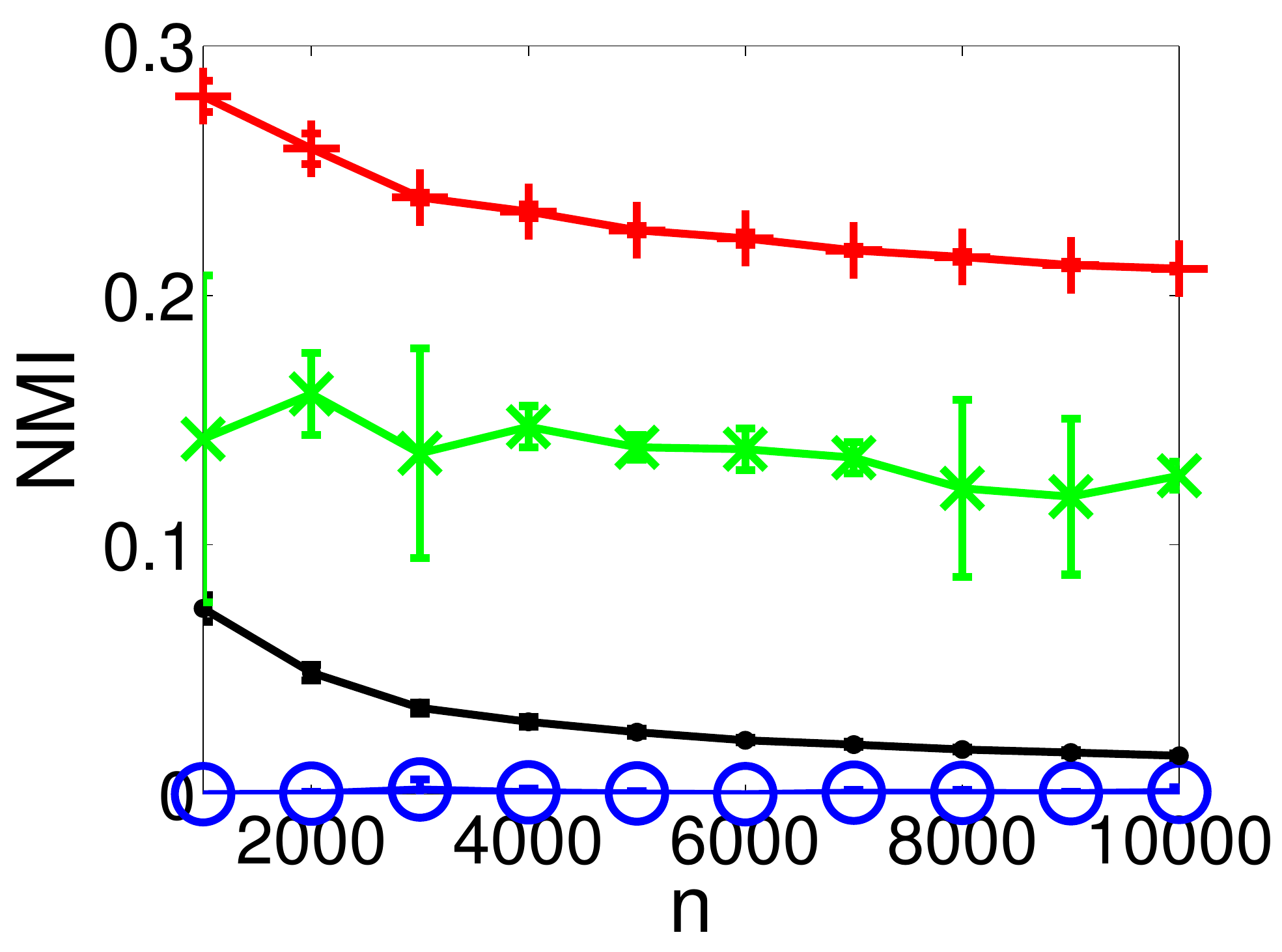} 
	\includegraphics[width=0.45\columnwidth]{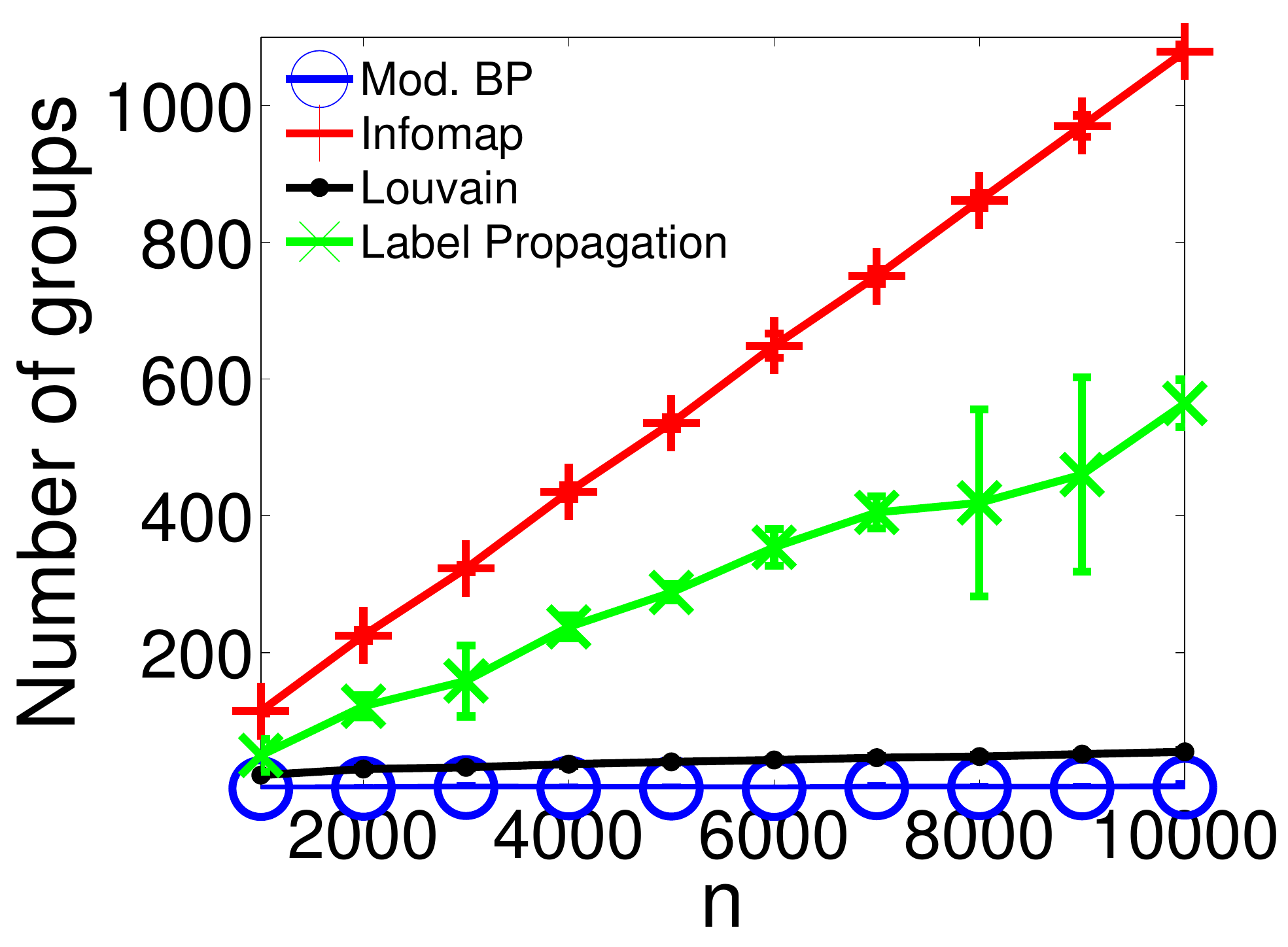}
	\caption{(Color online) Normalized Mutual Information (\textit{left}) 
	and number of groups (\textit{right}) given by 
	four algorithms on networks generated by Stochastic Block Model
	with $\epsilon=\frac{\pout}{\pin}=1$ (see text for the description of parameters of SBM), which is deep in the undetectable phase 
	of SBM with different system sizes. These algorithms are 
	(from top to bottom in left panel) Infomap \cite{Rosvall2008}, 
	Label Propagation \cite{raghavan2007near}, 
	Louvain method \cite{Blondel2008} and Modularity Belief Propagation \cite{Zhang2014pnas} .  
	These networks are essentially random graphs, 
	though in each network there is a planted partition, we expect no algorithm 
	can obtain any information about it. Each point in the figure is
	averaged over $10$ realizations.\label{fig:nmiran}
}
\end{figure}

After NMI was introduced as a metric for comparing community detection
algorithm, it becomes very popular in evaluating community detection algorithms. 
However in some cases we find that this metric gives un-consistent results.
One example is shown in Fig.~\ref{fig:nmiran} left where we compare
NMI between partitions obtained by four algorithms and the planted 
partition in the stochastic block model with parameter $\epsilon=1$. 
These four algorithms are Label Propagation \cite{raghavan2007near}, Infomap \cite{Rosvall2008}, 
Louvain method \cite{Blondel2008} and Modularity Belief Propagation \cite{Zhang2014pnas} respectively.
The principle of the algorithms are different: Label Propagation algorithm
maintains a group label for each node by iteratively adopting the label
that most of its neighbors have; Infomap method compresses a
description of information flow on the network; Louvain method maximizes
modularity by aggregation; and Modularity Belief Propagation detects 
a statistically significant community structures using the landscape
analysis in spin glass theory of statistical physics.

The stochastic block model (SBM) 
is also called the planted partition model. 
It has a planted, or ground-true, partition with $q$ groups of nodes, 
each node $i$ has a group label $t^*_i \in \{1, \ldots, q\}$. 
Edges in the network are generated independently according to 
a $q \times q$ matrix $p$, by connecting each pair of nodes $\brc{ij}$ 
with probability $p_{t^*_i,t^*_j}$. 
Here we consider the commonly studied case where the $q$ 
groups have equal size and where $p$ has only two distinct 
entries, $p_{rs}=\pin / n$ if $r=s$ and $\pout / n$ if $r \ne s$.  
We use $\eps=\pout/\pin$ to denote the ratio between these two entries, 
the larger $\eps$ the weaker the community structure is. With $\eps=0$
the network is essentially composed of two connected components, while
with $\eps=1$,
as in Fig.~\ref{fig:nmiran} left, the network is
deep in the un-detectable phase \cite{Decelle2011}, and are
essentially random graphs. In the later case though in each network 
there is a planted 
partition, the partition is not detectable in the sense that no algorithm 
could be able to find it or even find a partition 
that is correlated with it. This un-detectability of the planted partition 
in SBM has been proved for $q=2$ groups in \cite{Mossel12}.

However from Fig.~\ref{fig:nmiran} we can see that only Modularity BP gives zero NMI on
all networks that is consistent with the un-detectability we
described above, while other three algorithms give positive NMI, and Infomap 
gives a quite large NMI on all networks. Then the question arises: does it mean Infomap could do a theoretically-impossible job of 
finding the planted configuration in the un-detectable phase of SBM? We
think it is not the case, but the problem comes from NMI, 
the measure for how much one finds about the planted configuration.

In Fig.~\ref{fig:nmiran} right we plot the number of groups found by 
different algorithms, then we can see that only Modularity BP gives one group
while other algorithms report increasing number of groups when system 
size increases. Thus from this result we can guess that the large NMI 
found in Fig.~\ref{fig:nmiran} may 
come from a large number of group.

Recall that in computing NMI of two partitions, we use $\frac{n_a}{n}$ to 
approximate $p_a$, which is fine with $n\to\infty$ but leads to a finite size 
effect with $n$ finite. Since NMI can be seen as a function of entropies (as in Eq.~\eqref{eq:nmi}), we can 
express the finite size effect of NMI as the finite size effect of entropy,
which means that
entropy with an infinite system size $H_{\infty}(\bre{p_a})$ is different from entropy with a finite system size $\brc{H_{n}(\bre{n_a})}$,
where the expectation is taken over random instances.
Actually this effect comes from the fluctuations of $n_a$ around its mean
value $p_an$ and 
the concavity of entropy, as Jensen's inequality implies that
$$\brc{H_{n}(\bre{n_a})}\leq H_{\infty}(\bre{p_a}) = H_\infty(\bre{\brc{\frac{n_a}{n}}}).$$ 
More precisely, we have
\begin{align}
	\brc{H_{n}(\bre{n_a})}&=-\sum_{a=1}^q\brc{ \frac{n_a}{n}\log\frac{n_a}{n} }\nonumber\\
	&=-\sum_{a=1}^q\brc{ p_a\log p_a+\left.\frac{\partial (x\log x)}{\partial x} \right |_{x=p_a}(\frac{n_a}{n}-p_a)+
	\frac{1}{2}\left .\frac{\partial^2 (x\log x)}{\partial x^2} \right |_{x=p_a}(\frac{n_a}{n}-p_a)^2 }\nonumber.
\end{align}
Assuming further on the distribution that the random variable $n_a$ follows,
for example the Bernoulli distribution as in \cite{herzel1994finite}, the mean entropy at finite-size systems can be estimated by
inserting the first 
and second moment into last equation:
\begin{align}\label{eq:fsentropy}
	\brc{H_n(\bre{n_a})}
	&=-\sum_{a=1}^q\bra{p_a\log p_a +\frac{1-p_a}{2n}}\nonumber\\
	&=H_{\infty}(\bre{p_a})-\frac{q-1}{2n}.
\end{align}
Obviously using Eq.~\eqref{eq:I2} the finite size correction for mutual information is
$$
\brc{I_{\infty}(P_A,P_B)}-I_{n}(P_A,P_B)=\frac{q_a+q_b-q_aq_b-1}{2n}.
$$
In our system, if we treat number of nodes in a network as sample size, 
the difference between NMI estimated using \eqref{eq:nmi} in our network and 
in the network with an infinite number of nodes is expressed as
\begin{align}
	\brc{\NMI_{n}(P_A,P_B)}-\NMI_{\infty}(P_A,P_B)=\frac{1}{2n}\frac{q_aq_b-q_a-q_b+1}{H(P_A)+H(P_B)}.
\end{align}
One thing we can infer from last equation is that in finite-size networks, 
even two random partitions $A$ and $B$ have a non-vanishing NMI, and its value is
\begin{equation}\label{eq:nmiran}
	\NMI^{\rm{random}}_{n}(q_A,q_B)\approx\frac{1}{2n}\frac{q_aq_b-q_a-q_b+1}{H(P_A)+H(P_B)}.
\end{equation}
Note that in the last equation we put $\approx$ instead of $=$ because 
$\NMI^{\rm{random}}_{n}(q_A,q_B)$ represents NMI of an instance instead of 
the ensemble average.

\begin{figure}   
	\centering
	\includegraphics[width=0.55\columnwidth]{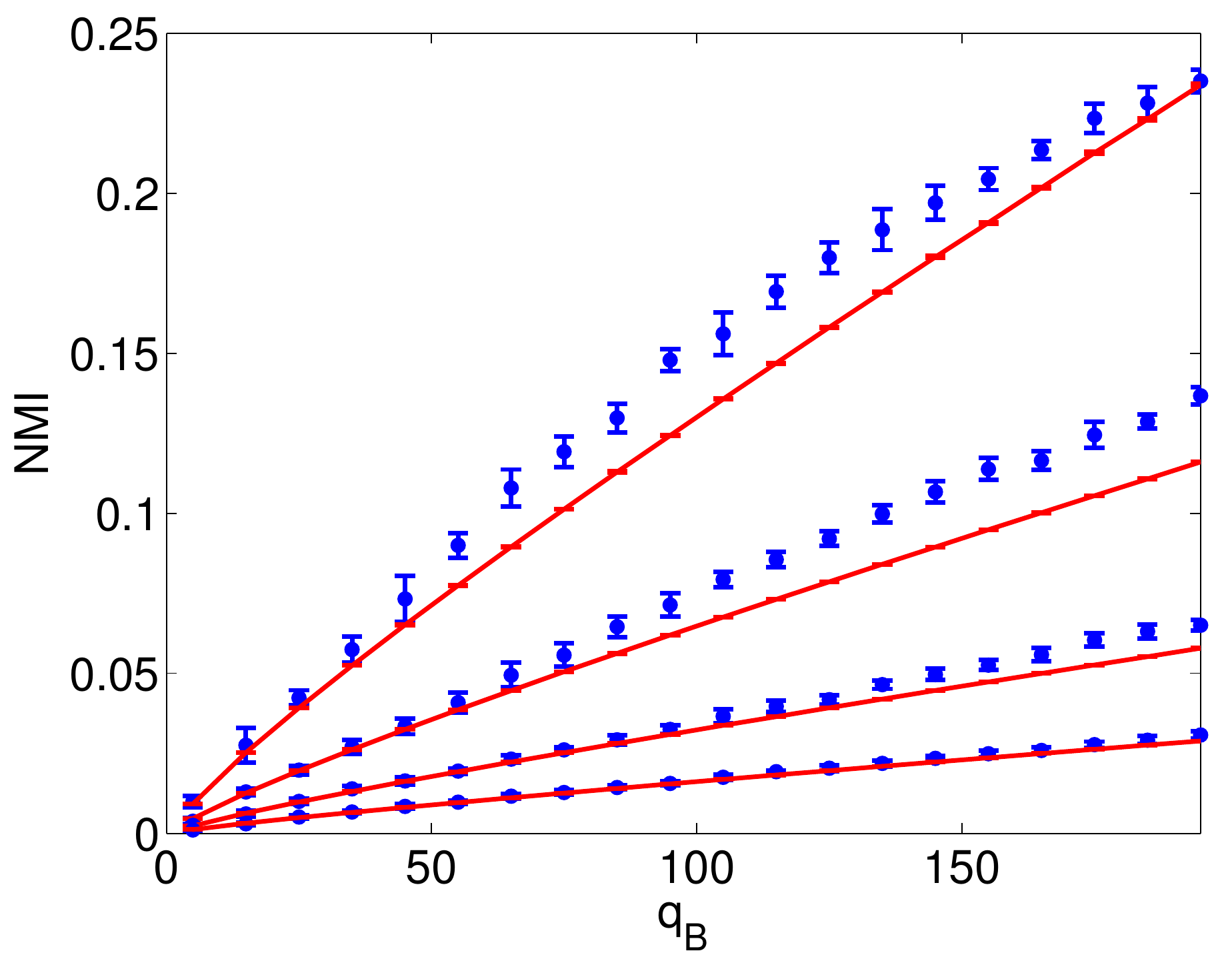}
	\caption{(Color online) Normalized Mutual Information between two random partitions $A$ and $B$.
	Partition $A$ always has $q_A=10$ groups, each of which has 
	expected size $\frac{1}{q_A}$. While $B$ has a varying number of groups $q_B$, with
	expected group size $\frac{1}{q_B}$. 
 In other words group labelling for each node is 
	chosen independently according to $\frac{1}{q_A}$ for partition $A$ 
	and $\frac{1}{q_B}$ for partition $B$. The lines are theoretical estimate 
	\eqref{eq:nmiran} and points with error-bars are experimental data, with each point 
	averaged over $10$ random instances. From top to bottom, system size are 
	$n=1000$,$2000$,$4000$ and $8000$.
\label{fig:nmitheory}
	}
\end{figure}

To test our theory on the finite-size correction, in Fig.~\ref{fig:nmitheory} we compare NMI of two random 
partitions $A$ and $B$ with 
$q_A=10$ groups in partition $A$ and a varying number of groups in partition $B$. We can 
see that Eq.~\eqref{eq:nmiran} gives a good estimate of NMI between
two random partitions. 
From the figure we see that 
for the same $n$, the finite-size correction is smaller with 
fewer states.
This is consistent with Eq.~\eqref{eq:nmiran} whose right hand side 
is an increasing function of $q_a$ and $q_b$.
Moreover we can see from the figure that
for the same $q$, the finite-size correction
is smaller with a larger system size, due to the $\frac{1}{n}$ dependence
in Eq.~\eqref{eq:nmiran}.
These two properties can also be used to understand the phenomenon we 
saw in Fig.~\ref{fig:nmiran} where NMI of Infomap and Label Propagation
change slowly with system size, because 
number of groups given by these two algorithms increases in system size. 
We note that in \cite{steuer2002mutual} a similar 
finite-size effect for Mutual Information has been studied, 
though in a different context.

From above analysis, we see that as a metric for accuracy of community
detection, NMI prefers a large 
number of partitions, which gives a systematic bias to evaluation results.
One way to fix this bias is to consider statistical significance of 
the NMI, by 
comparing it to NMI of a null model. In this article we choose 
a random configuration $C$, which 
has the same group-size distribution as the detected partition, as a null model, and 
define the \textit{relative Normalized Mutual Information} (rNMI) as
\begin{equation}\label{eq:rnmi}
	\rNMI(A,B)=\NMI(A,B)-\brc{\NMI(A,C)},
\end{equation}
where $\brc{\NMI(A,C)}$ is the expected NMI between 
the ground-true configuration $A$
and a random partition $C$, averaged over realizations of $C$.

So if partition $B$ has a large number of groups, though $\NMI(A,B)$ could be large,
a random configuration having same distribution as $B$ also has a large number 
of groups and hence a large $\NMI(A,C)$ and finally results to a small 
$\rNMI(A,B)$.

Actually the idea of computing statistical significance by comparing 
score of a partition to expected score of a null model has 
been used everywhere in 
science. For example, the well-known metric for community structures, 
Modularity \cite{Newman2004a}, compares the number of internal 
edges of a partition to the expected number of internal edges 
in random graphs which act as a null model.

An easy way to compute $\brc{\NMI(A,C)}$ is to generate several random 
configurations, 
compute NMI for each random configuration then take the average. In practice we find that usually $10$
realizations of $C$ are already enough. If we really care about 
the computational speed, we can use expression
$$\brc{\NMI(A,C)}=\frac{1}{2n}\frac{q_aq_b-q_a-q_b+1}{H(P_A)+H(P_B)}.$$ 
However as explained in Eq.~\eqref{eq:fsentropy}, this expression is only 
a first-order approximation adopting the Bernoulli distribution, hence 
is obviously less accurate as the simulation value, as shown
in Fig.~\ref{fig:nmitheory}.

\begin{figure}
   \centering
	\includegraphics[width=0.49\columnwidth]{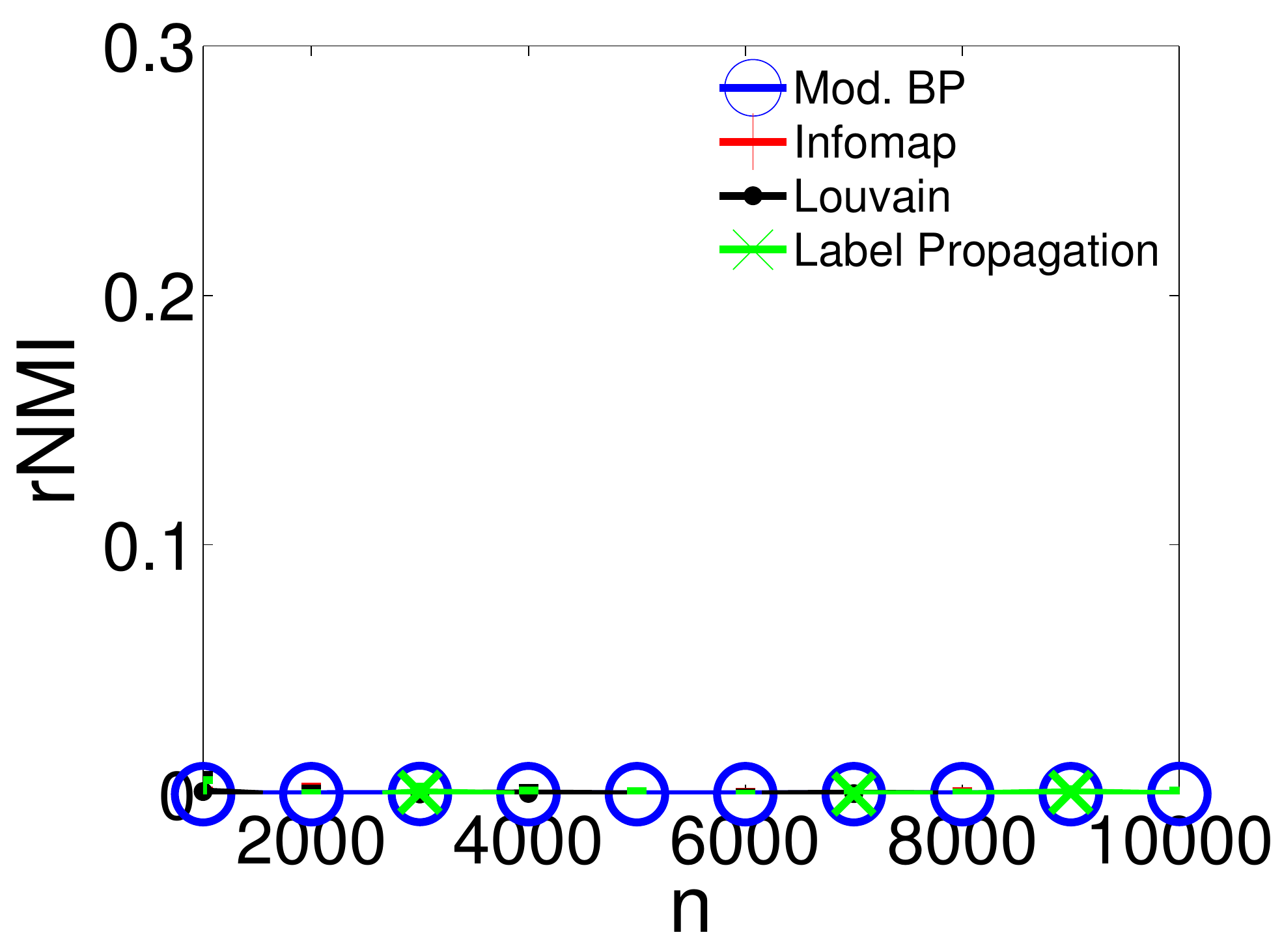}
	\includegraphics[width=0.45\columnwidth]{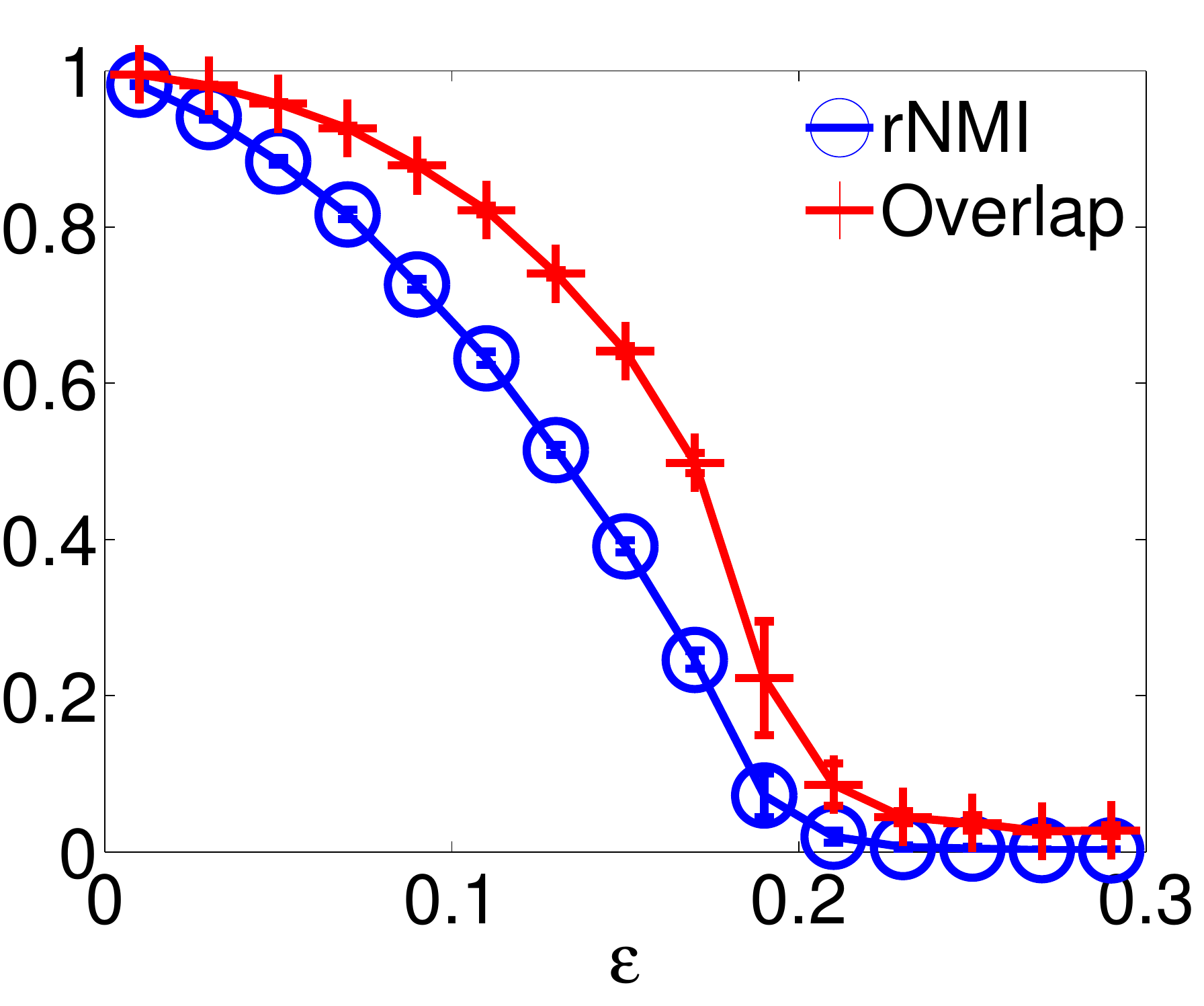}
	\caption{(Color online) (\textit{Left}) Relative Normalized Mutual Information given by the same four 
	algorithms on the same set of networks used in Fig~\ref{fig:nmiran}.
	Each point is averaged over $10$ realizations.\label{fig:rnmiran}
	(\textit{Right}) Relative Normalized Mutual Information compared with overlap 
	\eqref{eq:ovl} in networks generated by stochastic block model with $10000$ nodes,
	$6$ groups, average degree $6$ and different $\epsilon=\pout/\pin$.
	}
\end{figure}

In Fig.~\ref{fig:rnmiran} left we plot the rNMI given by the same four algorithms used 
in Fig.~\ref{fig:nmiran}, where we can see that now all algorithms 
report zero rNMI, telling us correctly that no one has found useful 
information about 
the planted partition of SBM networks in the un-detectable phase.

To test the accuracy of the proposed metric, in Fig.~\ref{fig:rnmiran} 
right we compare
the rNMI and overlap (Eq.~\eqref{eq:ovl}) between the planted partition
and the one detected by Belief Propagation (BP) algorithm, for benchmark
networks generated by stochastic block model. In this benchmark the 
detectability transition happens at $\epsilon^*=0.2$. It is known \cite{Decelle2011} 
that with $\epsilon < \epsilon ^*$ the planted configuration is detectable, and BP algorithm is
supposed to find a partition that is correlated with the planted configuration;
with 
$\epsilon > \epsilon ^*$ the planted configuration is un-detectable, 
which means that partition given by BP should be not correlated with 
the planted partition.
In this case $q_A=q_B$, so 
overlap defined in \eqref{eq:ovl} is a good metric and we can test whether rNMI 
gives the same information as overlap tells. In the figure we
see that the value of rNMI and overlap are consistent: they are both
high in detectable phase with $\epsilon<0.2$ and low in un-detectable phase with $\epsilon>0.2$.
Note that the overlap is not perfectly zero in un-detectable phase, 
because in maximizing overlap over permutations the effect of noise has 
been induced. So in this sense our metric which reports perfectly zero 
values in un-detectable phase, is a better metric for similarity of 
two partitions in undetectable phase than overlap, even when 
$q_A=q_B$.

\begin{figure}   \centering
	\includegraphics[width=0.45\columnwidth]{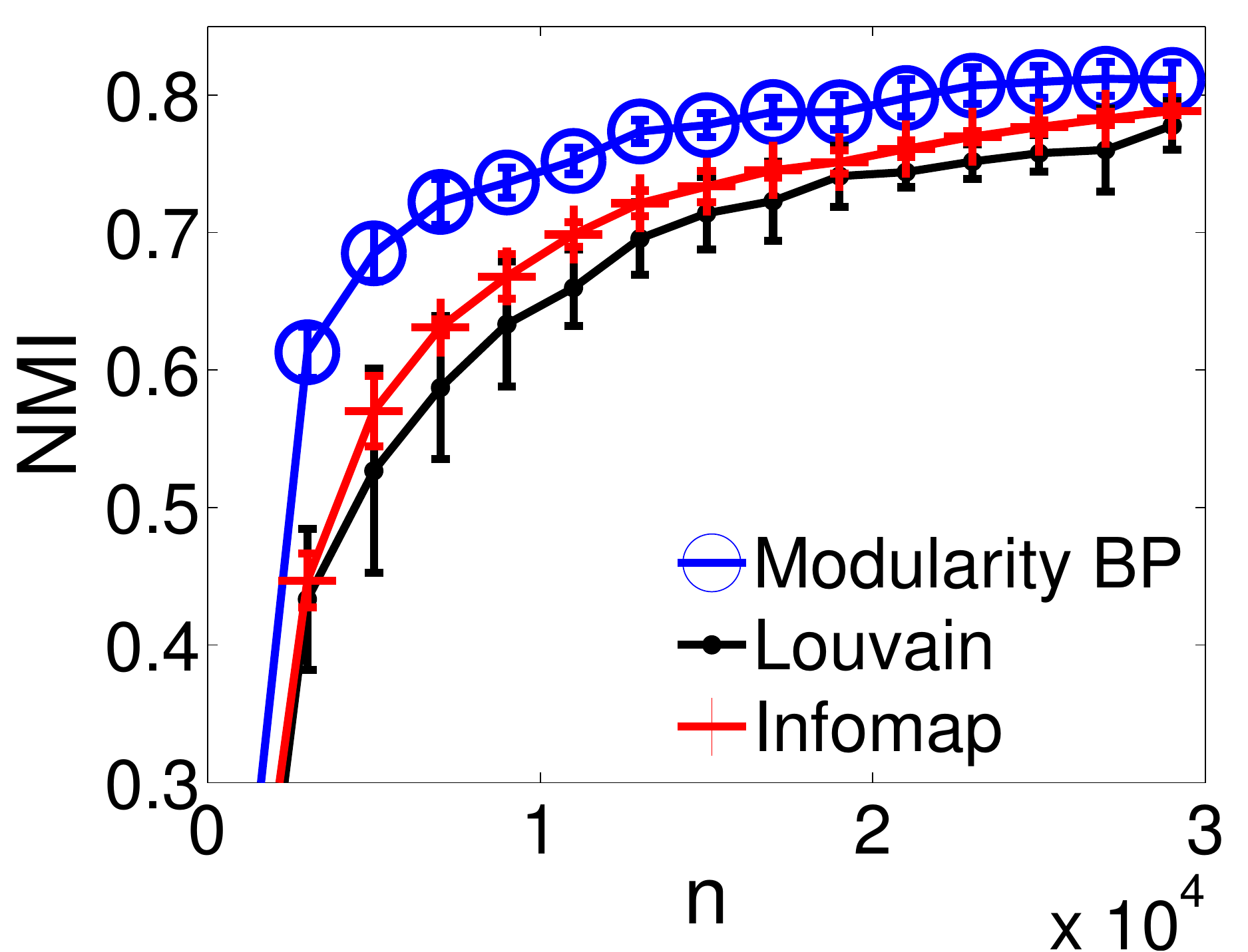}
	\includegraphics[width=0.45\columnwidth]{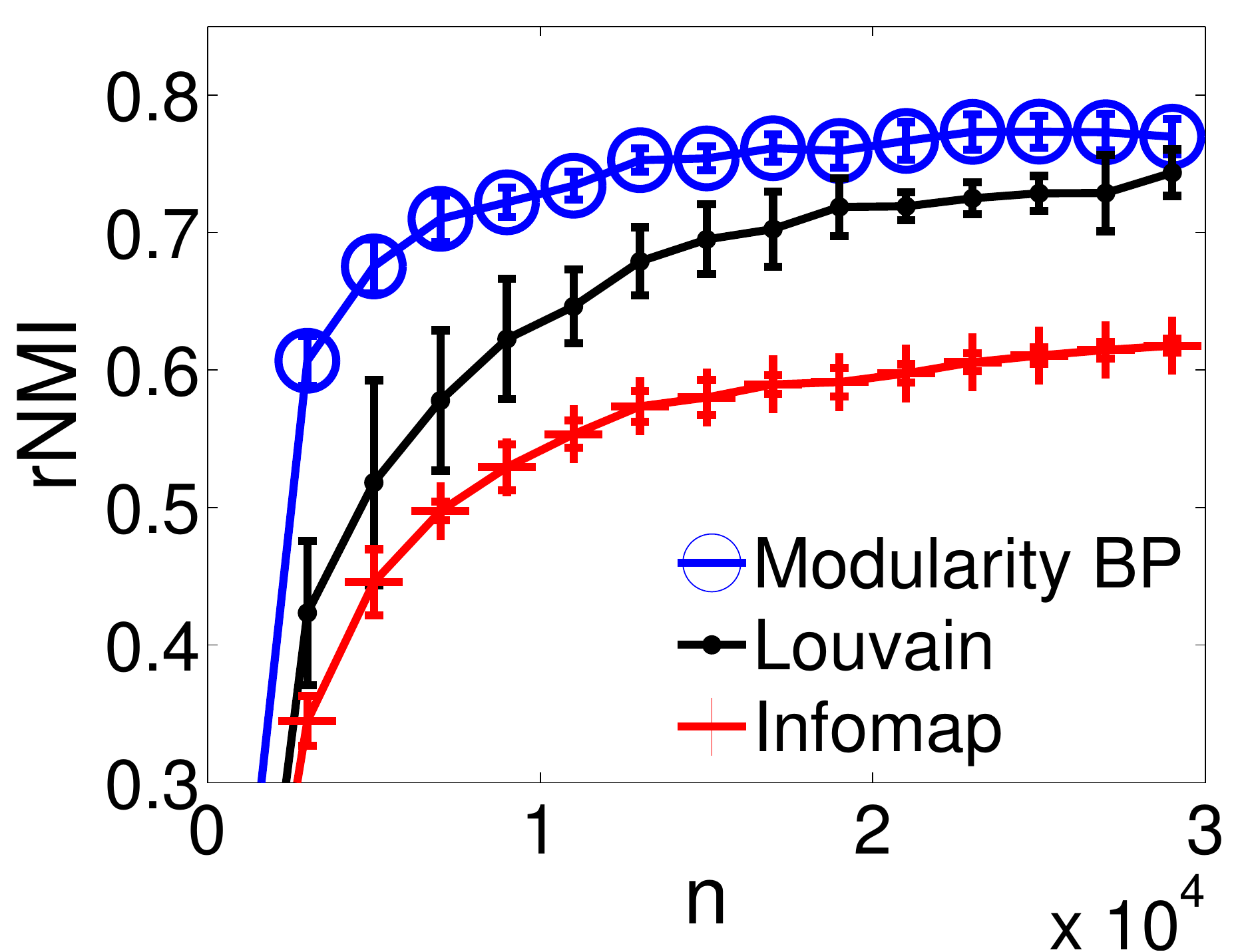}
	\caption{(Color online) Normalized Mutual Information (\textit{left}) and Relative Normalized Mutual Information (\textit{right}) 
	for three algorithms, Infomap \cite{Rosvall2008}, 
	Louvain~\cite{Blondel2008} and Modularity BP~\cite{Zhang2014pnas} 
	on LFR benchmarks~\cite{lancichinetti2008benchmark} with different 
	system sizes.
	The parameters of networks are: average degree $c=8$, 
	mixing parameter $\mu=0.45$, maximum degree is $50$, community sizes 
	range from $200$ to $400$, exponent of degree distribution is $-2$.
	and exponent of community size distribution is $-1$.\label{fig:lfr}
	}
\end{figure}

In Fig.~\ref{fig:lfr} we compare NMI and rNMI for three algorithms, 
Louvain, Infomap and 
Modularity BP, on benchmark networks generated 
by the LFR model \cite{lancichinetti2008benchmark} with different system sizes. 
the LFR model is also a planted model for generating benchmark networks 
with community structures. However compared with the SBM model, 
networks generated by the LFR model has a power-law 
degree distribution and group-size distribution.
From left panel of Fig.~\ref{fig:lfr} we can see that if 
we use the NMI as a metric for accuracy of these algorithms, 
we may conclude that 
Infomap works better than Louvain in whole set of benchmarks. However
from Fig.~\ref{fig:lfr} right we can see that Louvain actually works 
better than Infomap because it gives 
a larger rNMI than Infomap.
Moreover the difference of rNMI between Louvain and Infomap are larger 
with system size increases.
So Fig.~\ref{fig:lfr} also tells us that using NMI may give a wrong estimate of 
performance of community detection algorithms.

As a conclusion, in this article we showed analytically and numerically that using 
normalized mutual information as a metric for accuracy of community 
detection algorithms has a systematic error when number of groups given 
by algorithms are much different. We proposed to fix this problem 
by using the relative normalize mutual information which considers 
the statistically significance of NMI by comparing the NMI of two partitions
to the expected value of random partitions.
We note that there are other ways to estimate finite-size effect of entropy, e.g. a Bayesian estimate proposed in 
\cite{wolpert2013estimating}. We put it in future work to refine $\brc{\NMI(A,C)}$
in expression of rNMI \eqref{eq:rnmi} using Bayesian approaches.

Implementation of rNMI and examples of using it can be found at \cite{code}.

\begin{acknowledgements}
P.Z. was supported by Santa Fe Institute.
We are grateful to Cristopher Moore, Hyejin Youn and Sucheta Soundarajan
for helpful conversations. 
\end{acknowledgements}


\end{document}